\documentclass[twocolumn,fp]{jpsj3}
\usepackage{txfonts}
\usepackage{color}
\usepackage{bm}

\title{Electron Correlation Effects in Non-Centrosymmetric Metals \\ in the Weak Coupling Regime}

\author{Daisuke Maruyama$^{1}$ and Youichi Yanase$^{1,2,3}$\thanks{E-mail address: yanase@scphys.kyoto-u.ac.jp} %\\
}
\inst{\address{
$^{1}$Graduate School of Science and Technology, Niigata University, Niigata 950-2181, Japan 
\\ 
$^{2}$Department of Physics, Niigata University, Niigata 950-2181, Japan 
\\
$^{3}$Department of Physics, Graduate School of Science, Kyoto University, Kyoto 606-8502, Japan } 
}

\abst{
The two-dimensional Rashba-Hubbard model is investigated in order to clarify the electron correlation effects 
in non-centrosymmetric metals. 
The renormalization effect on Rashba spin-orbit coupling (RSOC) is calculated on the basis of 
second-order and third-order perturbation theories. 
We show that RSOC is enhanced by the electron correlation. 
On the other hand, the spin-splitting of the Fermi momentum (SFM) is robust against the electron correlation effect 
because the enhancement of RSOC is cancelled by the $k$-mass renormalization. 
The cancellation is almost perfect in often adopted models in which the momentum dependence of RSOC is 
linear in the quasiparticle velocity. A small correction to the SFM appears otherwise.  
We show that the same results are obtained for other antisymmetric spin-orbit couplings 
in non-centrosymmetric systems, such as the synthetic spin-orbit coupling in cold atoms.  
}

\begin{document}
\maketitle

\section{Introduction}

A breakthrough in exploring exotic states of matter induced by spin-orbit couplings has been one of the highlights 
of recent condensed matter physics. 
In particular, non-centrosymmetric metals lacking an inversion symmetry in the crystal structure exhibit 
various intriguing phenomena such as spintronics~\cite{Spin_Hall_1,Spin_Hall_2}, 
chiral and helical magnetism~\cite{Chiral_Magnetism_1,Chiral_Magnetism_2}, and non-centrosymmetric 
superconductivity~\cite{NCSC}. 
An antisymmetric spin-orbit coupling relating the momentum and spin of electrons appears owing to the 
lack of inversion symmetry, and it plays an essential role in these phenomena~\cite{NCSC}. 
Although various antisymmetric spin-orbit couplings exist depending on the symmetry of crystals~\cite{Frigeri_thesis},  
Rashba spin-orbit coupling (RSOC)~\cite{Rashba} has been investigated most intensively, probably because 
it appears not only in the bulk~\cite{PhysRevLett.92.027003,doi:10.1143/JPSJ.75.043703,PhysRevLett.95.247004,Ishizaka}
but also in artificial heterostructures~\cite{Ohtomo,Ueno_review,Shimozawa,Winkler_book}. 
The antisymmetric spin-orbit coupling gives rise to the spin-splitting of Fermi surfaces, and thus, 
a single-particle state acquires a spin texture in the momentum space~\cite{NCSC}. 
Such an electronic structure results in exotic quantum phases and quantum transport. 
Spin-split Fermi surfaces have been observed by angle-resolved photoemission spectroscopy 
(ARPES)~\cite{Ishizaka,Okuda_ARPES,Hirahara} 
and by de Haas-van Alphen (dHvA) measurement~\cite{Onuki}.

The interplay between the antisymmetric spin-orbit coupling and electron correlation effects often plays an important role. 
For instance, unconventional superconductivity with a giant upper critical field occurs 
in strongly correlated non-centrosymmetric systems~\cite{doi:10.1143/JPSJ.77.073705,PhysRevLett.98.197001,
PhysRevLett.101.267006}. 
Furthermore, it has been shown that novel electromagnetic responses, such as a magnetoelectric effect and 
anomalous Hall effect, are enhanced by the electron correlation~\cite{JPSJ.76.034712}. 
Thus, the interplay between the antisymmetric spin-orbit coupling and electron correlation appears to be 
an important issue. 
From the theoretical point of view, unconventional long-range order, 
such as magnetism~\cite{Yanase_CePt3Si,doi:10.1143/JPSJ.77.113706,JPSJ.77.124711} and 
superconductivity~\cite{Yanase_CePt3Si,JPSJ.77.124711,Tada_3,Takimoto_2,Yokoyama,PhysRevB.80.140509,PhysRevLett.101.267006,
PhysRevB.81.104506}, 
has been intensively clarified. 
Furthermore, spontaneous inversion-symmetry breaking and the emergence of an antisymmetric spin-orbit coupling 
due to electron correlation effects have been studied~\cite{doi:10.7566/JPSJ.83.014703,
PhysRevB.90.081115,Hayami_2,Hitomi,Hayami_3,Fu2015}. 
However, quasiparticles renormalized by electron correlation effects in the Fermi liquid state 
have only been investigated in a few works~\cite{JPSJ.76.034712,PhysRevB.81.104506,
PhysRevLett.101.267006}.

A Fermi liquid theory for non-centrosymmetric metals (chiral Fermi liquid theory) has been formulated on the basis of 
a two-dimensional Rashba-Hubbard model by Fujimoto~\cite{JPSJ.76.034712}, although only the diagonal self-energy 
was explicitly calculated on the basis of second-order perturbation theory. 
Although a phenomenological spin fluctuation model in three dimensions has been investigated by 
Tada {\it et al.}~\cite{PhysRevB.81.104506,PhysRevLett.101.267006}, again the off-diagonal self-energy 
was neglected. 
As shown in Ref.~\citen{JPSJ.76.034712}, the renormalization of spin-orbit coupling is neglected in these calculations. 
A one-dimensional Rashba-Hubbard model has been analyzed by Goth and Assaad~\cite{cond-mat.1406.7293}, but 
Landau quasiparticles break down in one-dimensional systems. 
In this paper, we clarify electron correlation effects in the chiral Fermi liquid state on the basis of 
the two-dimensional Rashba-Hubbard model by calculating both diagonal and off-diagonal self-energies 
on an equal footing. 
The calculation relies on the perturbation expansion with respect to the Coulomb interaction, and therefore, 
we obtain reliable results in the weak coupling regime. 

Two-dimensional electron gases in semiconductors have been theoretically studied  
for a similar purpose~\cite{Maslov2013}. 
Also, quasiparticle properties in the presence of antisymmetric spin-orbit coupling 
and a screened long-range Coulomb interaction have been investigated. It has been shown that the 
spin-orbit coupling is enhanced by the momentum dependence of the screened Coulomb interaction~\cite{Raikh1999}, 
but the quasiparticle properties are hardly affected by the interplay between them~\cite{Ross2005,Chesi2011,Chesi2012}. 
Interestingly we obtain similar results for correlated metals, although many-body effects arising from 
the short-range Coulomb interaction are the main subject of the paper. We also show the important role of 
anisotropy in the electron dispersion relation, which is not negligible in metals.

The paper is organized as follows.
In Sect.~\ref{FORM}, we introduce the Rashba-Hubbard model and formulate 
the renormalization of quasiparticles in the presence of electron correlation and spin-orbit coupling. 
Numerical and analytic results of second-order perturbation theory are shown in Sect.~\ref{2PT}.
In Sect.~\ref{3PT}, we show the results of third-order perturbation theory in order to clarify  
higher-order corrections. A brief summary and discussion are given in Sect.~\ref{SUM}.

\section{Formulation}\label{FORM}

In this section, we introduce the Rashba-Hubbard model and formulate the Green function, self-energy, 
and effective mass of quasiparticles in the presence of RSOC.

\subsection{Rashba-Hubbard model}

The effects of electron correlation in metals lacking an inversion symmetry are 
studied on the basis of the two-dimensional Rashba-Hubbard model, 
{\setlength\arraycolsep{1pt}
\begin{eqnarray}
H&=&H_0+H_{\rm int}, \\
H_0&=&\sum_{\mib{k},s}\varepsilon(\mib{k})c^{\dag}_{\mib{k}s}c_{\mib{k}s}+\alpha\sum_{\mib{k},s,s'}\mib{g}(\mib{k})\cdot\mib{\sigma}c^{\dag}_{\mib{k}s}c_{\mib{k}s'}, \\
H_{\rm int}&=&U\sum_in_{i\uparrow}n_{i\downarrow},
\end{eqnarray}
}
where $c_{\mib{k}s}$ ($c^{\dag}_{\mib{k}s}$) is the annihilation (creation) operator for an electron 
with momentum $\mib{k}$ and spin $s = \uparrow,\downarrow$. 
The electron number operator for spin $s$ at site $i$ is denoted as $n_{is}$.  

We consider a simple square lattice and assume a tight-binding model, 
$\varepsilon(\mib{k})=-2t_1(\cos{k_x}+\cos{k_y})+4t_2\cos{k_x}\cos{k_y}-\mu$, 
by taking account of the nearest-neighbour and next-nearest-neighbour hoppings. 
The chemical potential $\mu$ is involved in the dispersion relation. 
The second term in the single-particle part $H_0$ represents the antisymmetric spin-orbit coupling 
arising from the lack of inversion symmetry. 
Following the conventional notation~\cite{NCSC}, it is characterized by the g-vector $\mib{g}(\mib{k})$. 
We focus on RSOC, which has been studied in various fields of condensed matter physics. 
Thus, in Sects.~3 and 4 we will assume a g-vector that represents RSOC.  
Although the momentum dependence of the g-vector is determined by the orbital wave function 
in the Bloch state~\cite{JPSJ.77.124711}, we will adopt a simple form of RSOC, 
$\mib{g}(\mib{k})=2t_1(-\sin{k_y},\sin{k_x},0)$, which is justified in the absence of 
orbital degeneracy in the electronic structure~\cite{doi:10.7566/JPSJ.82.044711,doi:10.7566/JPSJ.82.083705}. 
On the other hand, we will discuss other kinds of antisymmetric spin-orbit coupling in Sect.~5. 
Thus, we adopt a general form of the g-vector in this section.

The interacting part $H_{\rm int}$ represents the on-site Coulomb repulsion. 
We investigate electron correlation effects in the weak coupling regime 
by using the perturbation expansion with respect to the Coulomb interaction $U$.

\subsection{Green function}

In this subsection, we formulate the renormalization of quasiparticles due to the electron correlation effect. 
In the presence of spin-orbit coupling, the noninteracting Green function is described 
in the matrix representation as 
{\setlength\arraycolsep{3pt}
\begin{eqnarray}
\hat{G}^{(0)}(\mib{k},{\rm i}\omega_{\rm n})&=&\left(
\begin{array}{ccc}
{\rm i}\omega_{\rm n}-\varepsilon-\alpha g_z & -\alpha(g_x-{\rm i}g_y) \\
-\alpha(g_x+{\rm i}g_y) &  {\rm i}\omega_{\rm n}-\varepsilon+\alpha g_z \\
\end{array}
\right)^{-1} \nonumber \\
&=&\sum_{\lambda=\pm}\left(\frac{\hat{I}+\lambda\frac{\mib{g}}{|\mib{g}|}\cdot\mib{\sigma}}{2}\right)
G^{(0)}_{\lambda}(\mib{k},{\rm i}\omega_{\rm n}), 
\label{nonGreen}
\end{eqnarray}
}
where $\hat{I}$ is the $2 \times 2$ unit matrix and $\omega_n=(2n-1)\pi T$ is the 
fermion Matsubara frequency. 
The noninteracting Green function in the chirality basis is obtained as
\begin{eqnarray}
G^{(0)}_{\lambda}(\mib{k},{\rm i}\omega_{\rm n})=\frac{1}{{\rm i}\omega_{\rm n}-\varepsilon(\mib{k})-\lambda\alpha|\mib{g}(\mib{k})|}, 
\label{nonband}
\end{eqnarray}
with $\lambda = \pm$ being the chirality index.

\begin{figure}[htbp]
  \begin{center}
    \includegraphics[scale=0.3]{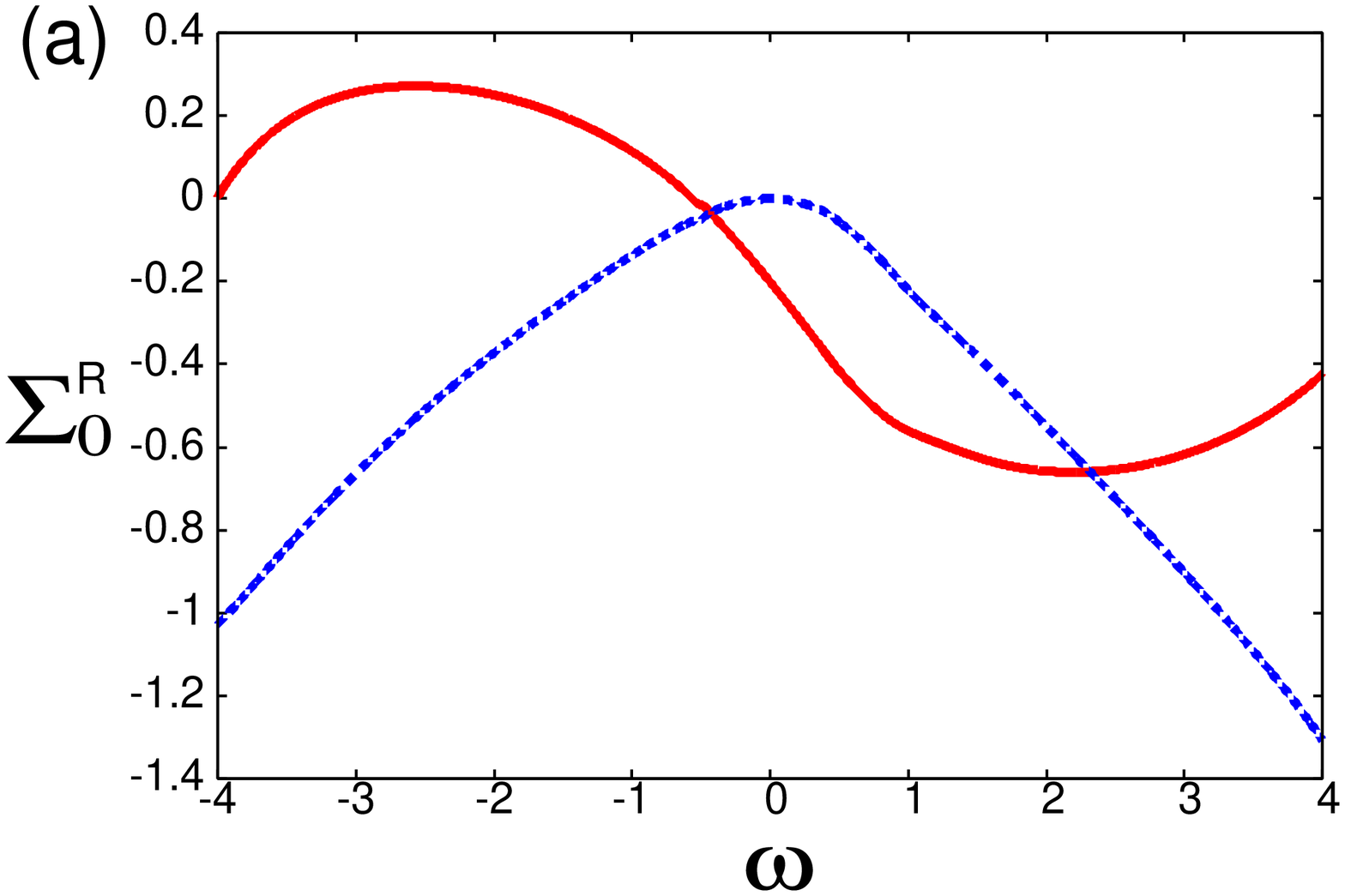}
    \includegraphics[scale=0.3]{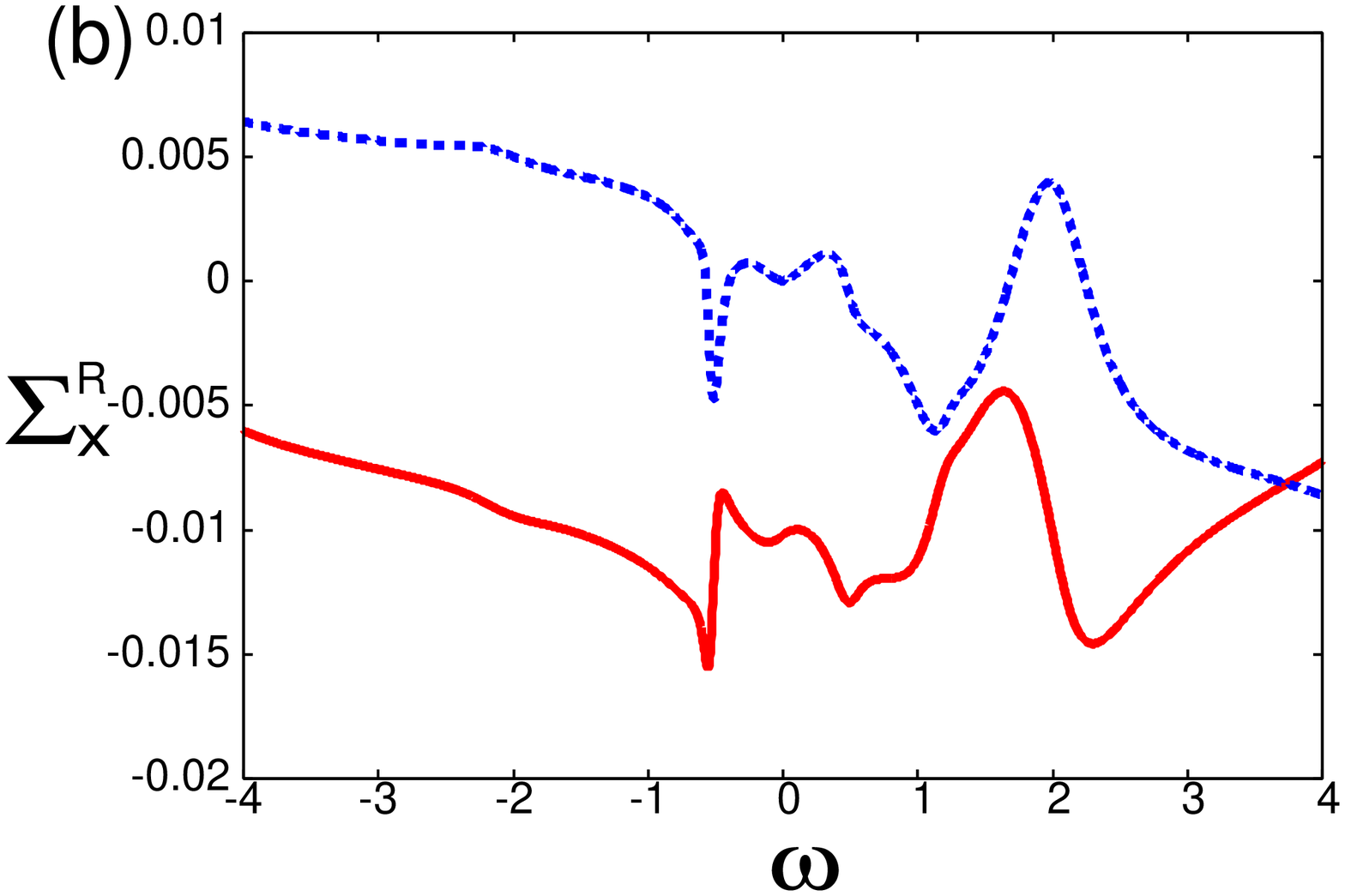}
    \includegraphics[scale=0.3]{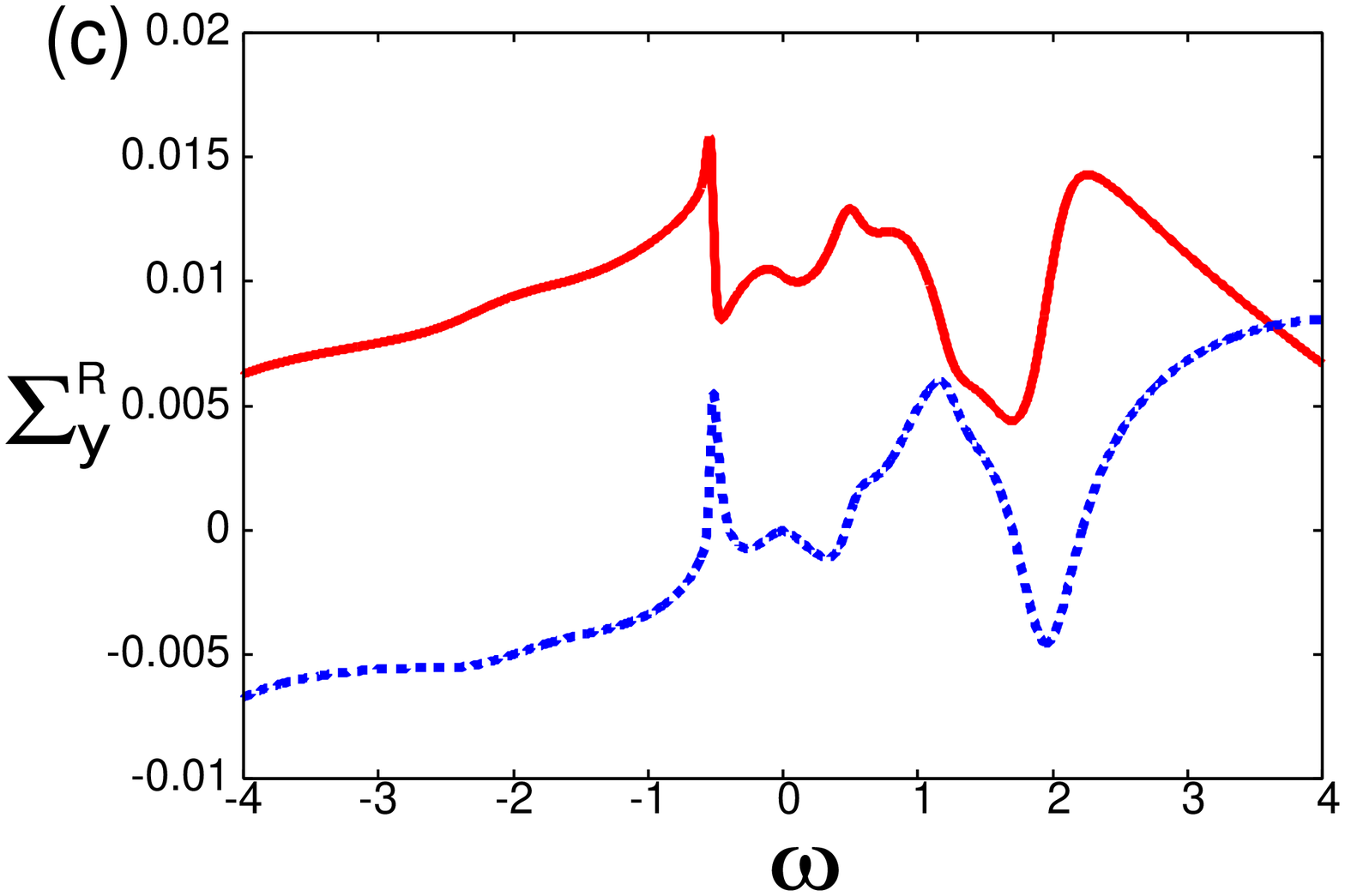}
  \end{center}
  \caption{(Color online)
    Frequency dependence of retarded self-energies at temperature $T=0.01$ 
    obtained by second-order perturbation theory. 
    We show the results at Fermi momenta $\mib{k} = \mib{k}_{\rm F+} \parallel$ [110] 
    for $(t_1, t_2)=(1,0)$, $U=4$, and $\alpha=0.1$. Solid (dashed) lines show the real (imaginary) part of 
    (a) $\Sigma_0^{\rm R}(k)$, (b) $\Sigma_x^{\rm R}(k)$, and (c) $\Sigma_y^{\rm R}(k)$.
    We adopted the P$\acute{\rm a}$de approximation for the analytic continuation.
  }
  \label{Figse}
\end{figure}

The dressed Green function is obtained by taking account of the self-energy 
$\hat{\Sigma}(k) = \left[\Sigma_{ss'}(k) \right]$. 
As we will show later, not only the diagonal self-energy $\Sigma_{\sigma\sigma}(k)$ 
but also the off-diagonal self-energy $\Sigma_{\sigma\bar{\sigma}}(k)$ 
plays an important role, although the latter was neglected in 
previous works~\cite{JPSJ.76.034712,PhysRevB.81.104506,PhysRevLett.101.267006}. 
Carrying out analytic continuation, the retarded Green function is described as 
{\setlength\arraycolsep{3pt}
\begin{eqnarray}
\hspace*{-2mm}
\hat{G}^{\rm R}(k)&=&\left(
\begin{array}{ccc}
\omega-\varepsilon-\alpha g_z-\Sigma^{\rm R}_{\uparrow\uparrow} & -\alpha(g_x-{\rm i}g_y)-\Sigma^{\rm R}_{\uparrow\downarrow} \\
-\alpha(g_x+{\rm i}g_y)-\Sigma^{\rm R}_{\downarrow\uparrow} &  \omega-\varepsilon+\alpha g_z-\Sigma^{\rm R}_{\downarrow\downarrow} \\
\end{array}
\right)^{-1}, 
\label{Green}
\end{eqnarray}
}
where $k=(\mib{k},\omega)$. 
Adopting the vector representation of the self-energy, 
\begin{eqnarray}
\hat{\Sigma}^{\rm R}(k) = \Sigma^{\rm R}_0(k) \, \hat{I} + \mib{\Sigma}^{\rm R}(k)\cdot\mib{\sigma},
\end{eqnarray} 
where $\mib{\Sigma}^{\rm R}=(\Sigma^{\rm R}_x,\Sigma^{\rm R}_y,\Sigma^{\rm R}_z)$,
$\Sigma^{\rm R}_0=(\Sigma^{\rm R}_{\uparrow\uparrow}+\Sigma^{\rm R}_{\downarrow\downarrow})/2$, 
$\Sigma^{\rm R}_x=(\Sigma^{\rm R}_{\downarrow\uparrow}+\Sigma^{\rm R}_{\uparrow\downarrow})/2$, 
$\Sigma^{\rm R}_y=(\Sigma^{\rm R}_{\downarrow\uparrow}-\Sigma^{\rm R}_{\uparrow\downarrow})/2{\rm i}$, 
and $\Sigma^{\rm R}_z=(\Sigma^{\rm R}_{\uparrow\uparrow}-\Sigma^{\rm R}_{\downarrow\downarrow})/2$, 
the dressed Green function is represented in a familiar form, 
{\setlength\arraycolsep{3pt}
\begin{eqnarray}
\hat{G}^{\rm R}(k)
&=&\left(
\begin{array}{ccc}
\omega-\varepsilon'-\alpha g_z' & -\alpha(g_x'-{\rm i}g_y') \\
-\alpha(g_x'+{\rm i}g_y') &  \omega-\varepsilon'+\alpha g_z' \\
\end{array}
\right)^{-1} \nonumber \\
&=&\sum_{\lambda=\pm}\left(\frac{\hat{I}+\lambda\frac{\mib{g}'}{|\mib{g}'|}\cdot\mib{\sigma}}{2}\right)G^{\rm R}_{\lambda}(k). 
\label{reGreen}
\end{eqnarray}}
The renormalized Green function in the chirality basis is obtained as 
\begin{eqnarray}
G^{\rm R}_{\lambda}(k)=\frac{1}{\omega-\varepsilon'(k)-\lambda\alpha|\mib{g}'(k)|}, 
\label{band}
\end{eqnarray}
where $\varepsilon'(k)\equiv\varepsilon(\mib{k})+\Sigma^{\rm R}_0(k)$. 
The renormalization of spin-orbit coupling is taken into account by the self-energy correction of the g-vector,  
$\alpha \mib{g}'(k) \equiv \alpha \mib{g}(\mib{k}) + {\rm Re} \mib{\Sigma}^{\rm R}(k)$~\cite{comment1}. 
Here we dropped the imaginary part of the spin-dependent self-energy Im$\mib{\Sigma}^{\rm R}(k)$, because 
it is negligible at low temperatures in a Fermi liquid state, Im$\mib{\Sigma}^{\rm R}(k) \propto T^2$. 
Figure~\ref{Figse} shows the self-energy obtained by second-order perturbation theory, 
and we indeed see the negligible imaginary part around $\omega=0$. 
This property is not altered in third-order perturbation theory, indicating the Fermi liquid state.

The Fermi surfaces of spin-split bands are defined by the singularity of the Green function and are thus 
obtained by solving the equation
\begin{eqnarray}
\varepsilon'(\mib{k}_{{\rm F}\lambda},0)+\lambda\alpha|\mib{g}'(\mib{k}_{{\rm F}\lambda},0)|=0.
\end{eqnarray}
The Fermi momentum of the $\lambda$-band is denoted as $\mib{k}_{{\rm F}\lambda}$ and 
the spin-splitting of the Fermi momentum (SFM) is defined as $\Delta k_{\rm F}=|\mib{k}_{{\rm F}+}-\mib{k}_{{\rm F}-}|$.

We also take into account the correlation correction of the chemical potential. 
First, we calculate the chemical potential $\mu_0$ at $U=0$ for which the electron density per site is $n$. 
Next, we calculate the self-energy, and the chemical potential is corrected 
as $\mu = \mu_0 + \delta\mu$, for which the dressed Green function leads to the electron density, 
lim$_{\eta \rightarrow +0}{\rm Tr} \sum_{k} \hat{G}(k) e^{i \omega_n \eta} 
= {\rm Tr} \sum_{k} \hat{G}(k) + 1 = n$.

\subsection{$k$-mass and $\omega$-mass} 

Next, we introduce the effective mass of quasiparticles. 
The effective mass of the $\lambda$-band $m_{\lambda}^{\ast}$ is obtained as the product of the $\omega$-mass and $k$-mass,
{\setlength\arraycolsep{1pt}
\begin{eqnarray}
\frac{m_{\lambda}^{\ast}}{m_{\lambda}}
%\biggl{|}_{\substack{\mib{k}=\mib{k}_{{\rm F}\lambda} \\ \omega=0}}
=\frac{m_{\lambda}^{\omega}}{m_{\lambda}}
%\biggl{|}_{\substack{\mib{k}=\mib{k}_{{\rm F}\lambda} \\ \omega=0}}
\times\frac{m_{\lambda}^{{\rm k}}}{m_{\lambda}}, 
%\biggl{|}_{\substack{\mib{k}=\mib{k}_{{\rm F}\lambda} \\ \omega=0}},
\label{effectivemass}
\end{eqnarray}}
where $m_{\lambda}$ is the bare mass of the $\lambda$-band. 
The $\omega$-mass is given by the frequency derivative of the self-energy, 
{\setlength\arraycolsep{1pt}
\begin{eqnarray}
\frac{m_{\lambda}^{\omega}}{m_{\lambda}}
%\biggl{|}_{\substack{\mib{k}=\mib{k}_{{\rm F}\lambda} \\ \omega=0}}
=1-\frac{\partial{\rm Re}\Sigma^{\rm R}_0(\mib{k}_{{\rm F}\lambda},\omega)}{\partial\omega}\biggl{|}_{\omega=0}-\lambda\alpha\frac{\partial|\mib{g}'(\mib{k}_{{\rm F}\lambda},\omega)|}{\partial\omega}\biggl{|}_{\omega=0}.
\label{omegamass}
\end{eqnarray}}
Furthermore, quasiparticles acquire a $k$-mass renormalization through the 
momentum derivative of the self-energy, 
{\setlength\arraycolsep{1pt}
\begin{eqnarray}
\frac{m_{\lambda}^{\rm k}}{m_{\lambda}}
%\biggl{|}_{\substack{\mib{k}=\mib{k}_{{\rm F}\lambda} \\ \omega=0}}
&=&\frac{\frac{\partial\varepsilon(\mib{k})}{\partial\mib{k}}+\lambda\alpha\frac{\partial|\mib{g}(\mib{k})|}{\partial\mib{k}}}{\frac{\partial\varepsilon(\mib{k})}{\partial\mib{k}}+\frac{\partial{\rm Re}\Sigma^{\rm R}_0(\mib{k},0)}{\partial\mib{k}}+\lambda\alpha\frac{\partial|\mib{g}'(\mib{k},0)|}{\partial\mib{k}}}\Biggl{|}_{\mib{k}=\mib{k}_{{\rm F}\lambda}}.
\label{kmass}
\end{eqnarray}}
The $k$-mass renormalization is often neglected in Fermi liquid theory for strongly correlated 
electron systems because it is quantitatively less important than the $\omega$-mass renormalization. 
However, the $k$-mass renormalization plays an essential role in correlated non-centrosymmetric metals 
as we will show below.

We adopt an approximation formula for a numerical calculation. 
We obtain the self-energy at $\omega=0$ through the Matsubara self-energy 
{\setlength\arraycolsep{1pt}
\begin{eqnarray}
\Sigma^{\rm R}_{\alpha}(\mib{k},\omega=0)\simeq\frac{\Sigma_{\alpha}(\mib{k},\omega_{\rm n}=\pi T)+\Sigma_{\alpha}(\mib{k},\omega_{\rm n}=-\pi T)}{2}.
\label{oemga-mass_numerical}
\end{eqnarray}} 
The temperature is assumed to be $T=0.01$ in the following numerical results. 
Frequency derivatives are calculated by using the Kramers-Kronig relation as  
{\setlength\arraycolsep{1pt}
\begin{eqnarray}
\hspace*{0mm}
\frac{\partial{\rm Re}\Sigma^{\rm R}_{\alpha}(\mib{k},\omega)}{\partial\omega}\biggl{|}_{\omega=0}
&=&\frac{\partial{\rm Im}\Sigma_{\alpha}(\mib{k},\omega_{\rm n})}{\partial\omega_{\rm n}}\biggl{|}_{\omega_{\rm n}=0} \nonumber \\
&\simeq&\frac{{\rm Im}\Sigma_{\alpha}(\mib{k},\omega_{\rm n}=\pi T)-{\rm Im}\Sigma_{\alpha}(\mib{k},\omega_{\rm n}=-\pi T)}{2\pi T}. 
\nonumber \\
\label{k-mass_numerical}
\end{eqnarray}}
We also calculated the retarded self-energy $\Sigma^{\rm R}_\alpha(k)$
using the P$\acute{\rm a}$de approximation (see Fig.~\ref{Figse}) 
and estimated the $\omega$-mass and $k$-mass. 
It has been confirmed that the two numerical estimations coincide with each other. 
Thus, the approximation formulas Eqs.~(\ref{oemga-mass_numerical}) and (\ref{k-mass_numerical}) have been justified.

\section{Second-Order Perturbation Theory}\label{2PT}

\begin{figure}[htbp]
  \begin{center}
    \includegraphics[scale=0.4]{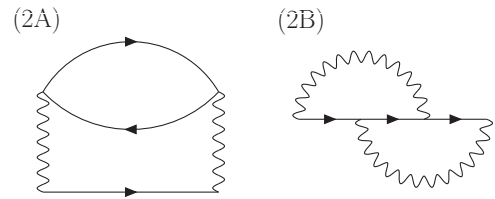}
  \end{center}
  \caption{Skeleton diagrams of self-energy in second-order perturbation theory. 
    The solid and wavy lines show the bare Green function $G^{(0)}_{ss'}(k)$ and Coulomb interaction $U$, respectively.
    }
  \label{Figse2o}
\end{figure}

We here show the results of second-order perturbation theory for the two-dimensional Rashba-Hubbard model. 
Since the first-order self-energy is involved in the correction of the chemical potential, second-order 
perturbation theory is the lowest-order theory justified in the weak coupling regime. 
We calculate the second-order self-energy represented by the skeleton diagrams in Fig.~\ref{Figse2o}. 
In the absence of RSOC, the self-energy corrections represented by Fig.~\ref{Figse2o}(2B) disappear. 
On the other hand, these terms give rise to the renormalization of RSOC. 
Thus, we take into account both Figs.~\ref{Figse2o}(2A) and \ref{Figse2o}(2B), 
although the latter was neglected in a previous study~\cite{JPSJ.76.034712}.

\subsection{Numerical results}\label{NUMERES}

First, we show the numerical results of the effective mass, the renormalization of RSOC, and the SFM. 
We choose the coupling constant of RSOC as $\alpha=0.1$ throughout the paper. 
We assume $U=4$ unless otherwise specified. 
In this subsection, the next-nearest-neighbour hopping is neglected for simplicity, and thus we assume $(t_1,t_2)=(1,0)$. 
Then, the Rashba-type g-vector is represented by the velocity as 
$\mib{g}(\mib{k})=2t_1(-\sin{k_y},\sin{k_x},0)= [-v_y(\mib{k}),v_x(\mib{k}),0]$. 
In the next subsection (Sect.~3.2), we will show that this relation between the RSOC and the velocity plays an essential role.

\begin{figure}[htbp]
\begin{center}
    \includegraphics[scale=0.33]{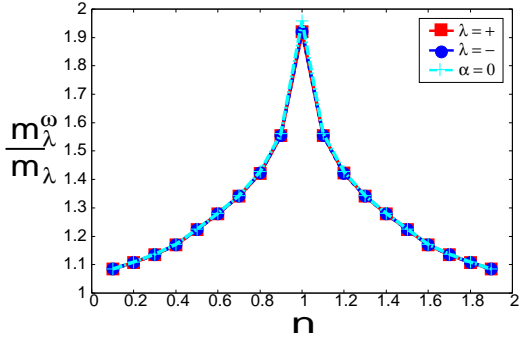}
  \end{center}
  \caption{(Color online)
    Filling dependence of the $\omega$-mass at $\mib{k} = \mib{k}_{{\rm F} \lambda} \parallel$ [110] 
    obtained by second-order perturbation theory. 
    Squares and circles show the $\omega$-mass of the $\lambda = +$ and $-$ band, respectively. 
    The $\omega$-mass in the absence of RSOC ($\alpha=0$) is shown by crosses. 
  }
  \label{Figom}
\end{figure}

Figure~\ref{Figom} shows the $\omega$-mass as a function of the electron filling $n$. 
As is known from Fermi liquid theory, the effective mass is enhanced by the electron correlation 
through the $\omega$-mass. 
The mass enhancement is pronounced near the half-filling, $n=1$, because of the large density of states (DOS) 
due to the Van-Hove singularity at $\mib{k} = (\pi,0)$ and $(0,\pi)$.  
Because we assume a small RSOC compared with the Fermi energy, the effect of RSOC on the $\omega$-mass is negligible. 
Indeed, the three lines in Fig.~\ref{Figom} almost coincide with each other. Thus, the band dependence of the effective mass 
is small in the weak coupling regime.

\begin{figure}[htbp]
\begin{center}
    \includegraphics[scale=0.33]{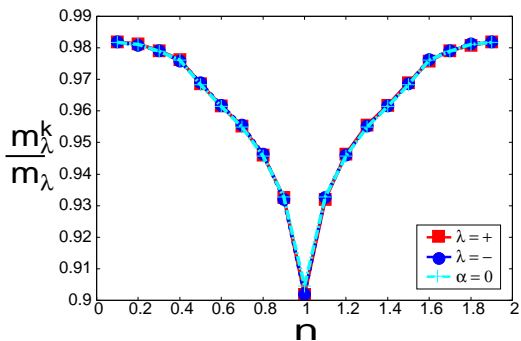}
  \end{center}
  \caption{(Color online)
    Filling dependence of the $k$-mass at $\mib{k} = \mib{k}_{{\rm F \lambda}} \parallel$ [110] 
    obtained by second-order perturbation theory. 
    Squares and circles show the $k$-mass of the $\lambda = +$ and $-$ band, respectively, while 
    crosses show the $k$-mass in the absence of RSOC ($\alpha=0$). 
  }
  \label{Figkm}
\end{figure}

In contrast to the $\omega$-mass, the $k$-mass is suppressed by electron correlation effects. 
Indeed, Fig.~\ref{Figkm} shows that $m_{\lambda}^{{\rm k}}/m_{\lambda} <1$. 
However, the total effective mass [Eq.~(\ref{effectivemass})] is enhanced because the $\omega$-mass renormalization 
is much larger than the $k$-mass renormalization. 
For $|\alpha| \ll 1$, both $\omega$-mass and $k$-mass renormalization mainly originate from the diagonal component 
of self-energy $\Sigma_0^{(2A)}(k)$ represented by Fig.~\ref{Figse2o}(2A).

\begin{figure}[htbp]
  \begin{center}
    \includegraphics[scale=0.35]{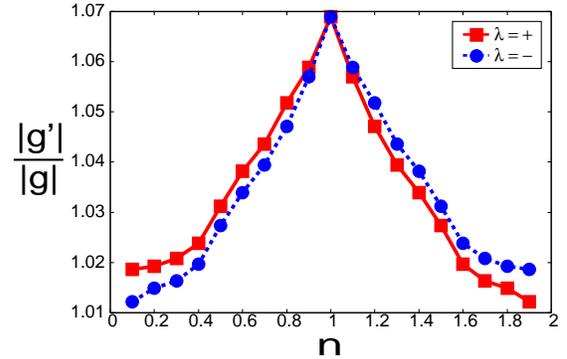}
  \end{center}
  \caption{(Color online)
    Filling dependence of the renormalization of RSOC obtained by second-order perturbation theory. 
    Squares and circles show $|\mib{g}'(\mib{k},\omega=0)|/|\mib{g}(\mib{k})|$ at the Fermi momentum 
    $\mib{k} = \mib{k}_{{\rm F} \lambda} \parallel$ [110] for $\lambda=+$ and $\lambda=-$, respectively. 
  }
  \label{Figrg}
\end{figure}

Now we discuss the renormalization of RSOC due to the electron correlation effect, 
which is given by $|\mib{g}'(\mib{k},\omega=0)|/|\mib{g}(\mib{k})|$.  
Figure~\ref{Figrg} shows that $|\mib{g}'(\mib{k},\omega=0)|/|\mib{g}(\mib{k})| > 1$ around the Fermi surface 
irrespective of the electron density, and thus RSOC is enhanced by the electron correlation  
through the spin-dependent part of the self-energy $\mib{\Sigma}(k)$. 
Because the main contribution to $\mib{\Sigma}(k)$ comes from the diagram represented in Fig.~\ref{Figse2o}(2B), 
it is essential to take into account these terms when studying of correlation effects on the RSOC. 

For our choice of hopping integrals $(t_1,t_2)=(1,0)$, the Hamiltonian has a particle-hole 
symmetry because $\varepsilon(\mib{k}) + \mu= - \varepsilon(\mib{Q}-\mib{k}) -\mu$ and 
$\mib{g}(\mib{Q}-\mib{k}) = \mib{g}(\mib{k})$ with $\mib{Q}= (\pi,\pi)$. 
Then, the particle-hole transformation, $c_{\mib{k}s} \rightarrow c_{\mib{Q}-\mib{k}s}^{\dag}$, changes the signs of 
the chemical potential ($\mu \rightarrow -\mu$) and RSOC ($\alpha \rightarrow -\alpha$), 
and thus the electron density is changed as $n \rightarrow 2-n$. 
Therefore, the renormalizations of the $\omega$-mass, $k$-mass, and RSOC show symmetric behaviors 
$\frac{m_{\lambda}^{\omega,\,\,k}}{m_{\lambda}}\mid_{\,n=n_0} = \frac{m_{-\lambda}^{\omega,\,\,k}}{m_{-\lambda}}\mid_{\,n=2-n_0}$ and
$|\mib{g}'(\mib{k}_{{\rm F} \lambda}, 0)|/|\mib{g}(\mib{k}_{{\rm F} \lambda})|_{\,n=n_0}= 
|\mib{g}'(\mib{k}_{{\rm F} -\lambda}, 0)|/|\mib{g}(\mib{k}_{{\rm F} -\lambda})|_{\,n=2-n_0}$ 
(see Figs.~3-5).

\begin{figure}[htbp]
  \begin{center}
    \includegraphics[scale=0.42]{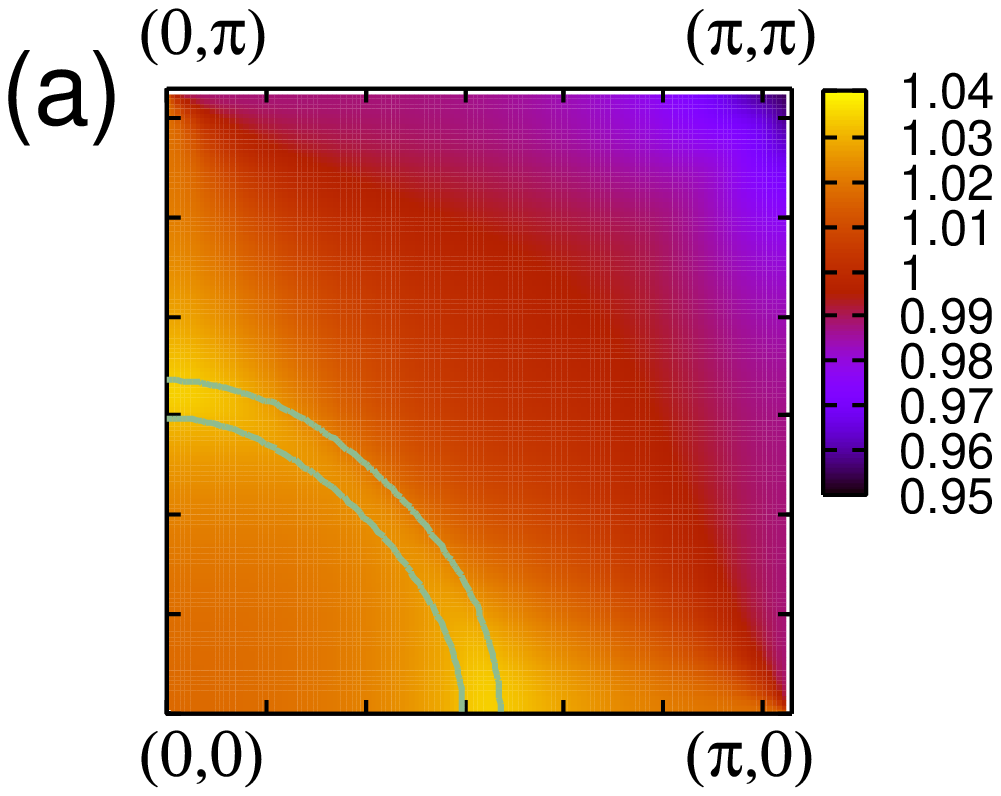}
    \includegraphics[scale=0.42]{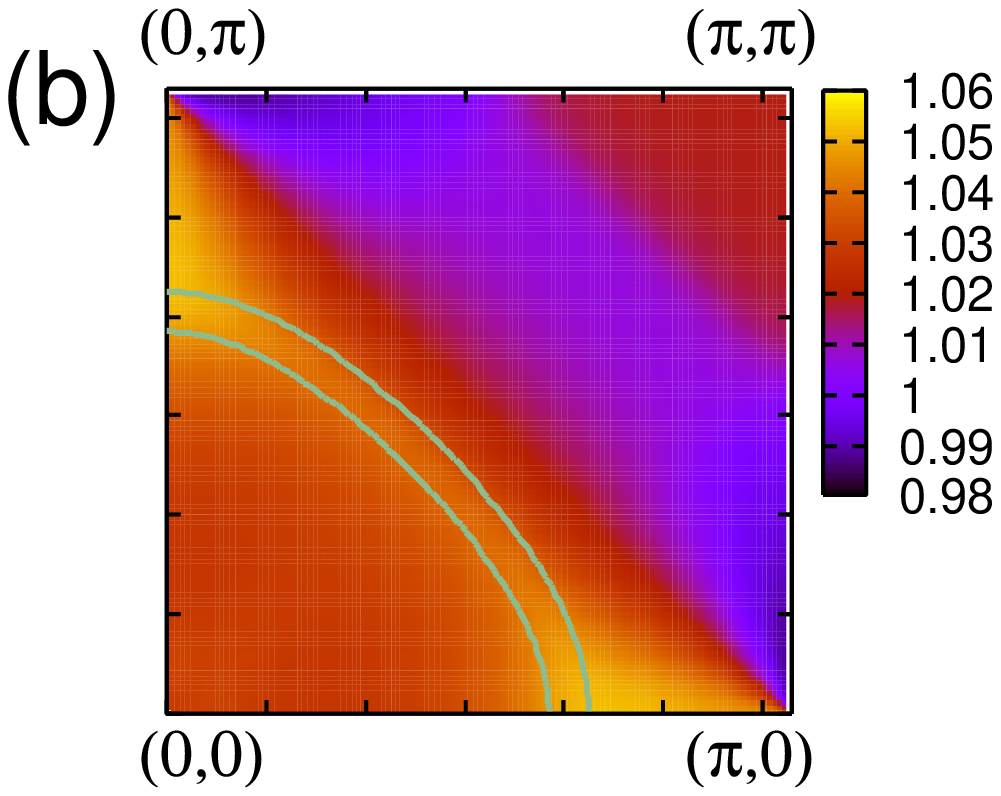}
\\
    \includegraphics[scale=0.42]{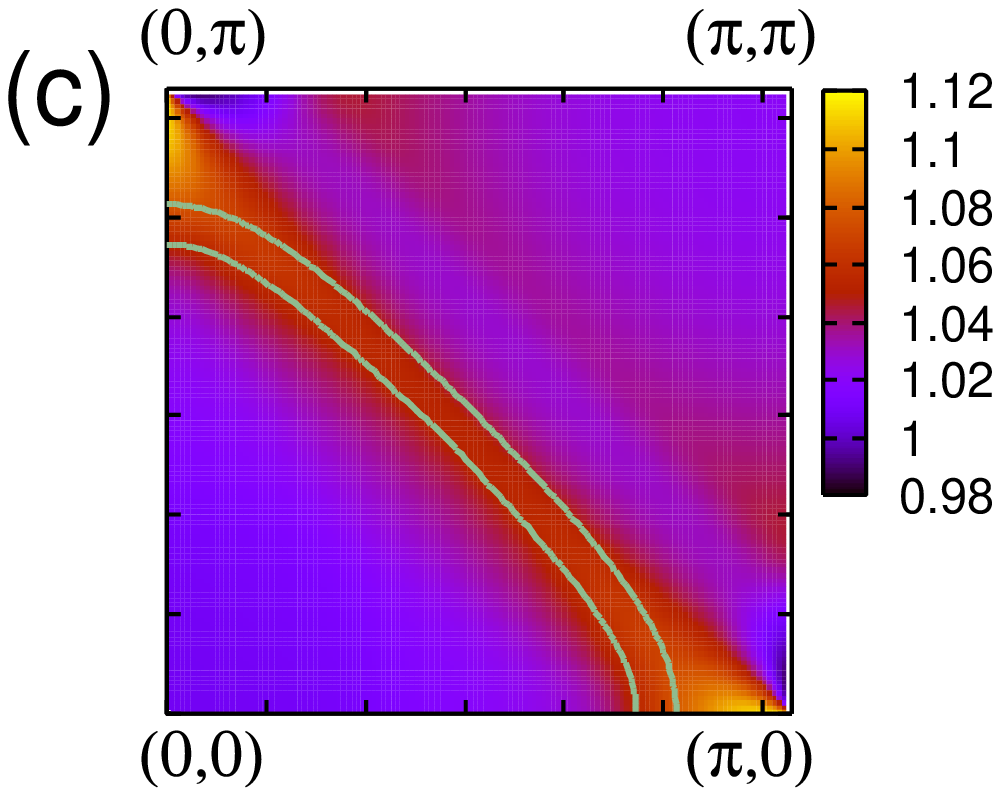}
    \includegraphics[scale=0.42]{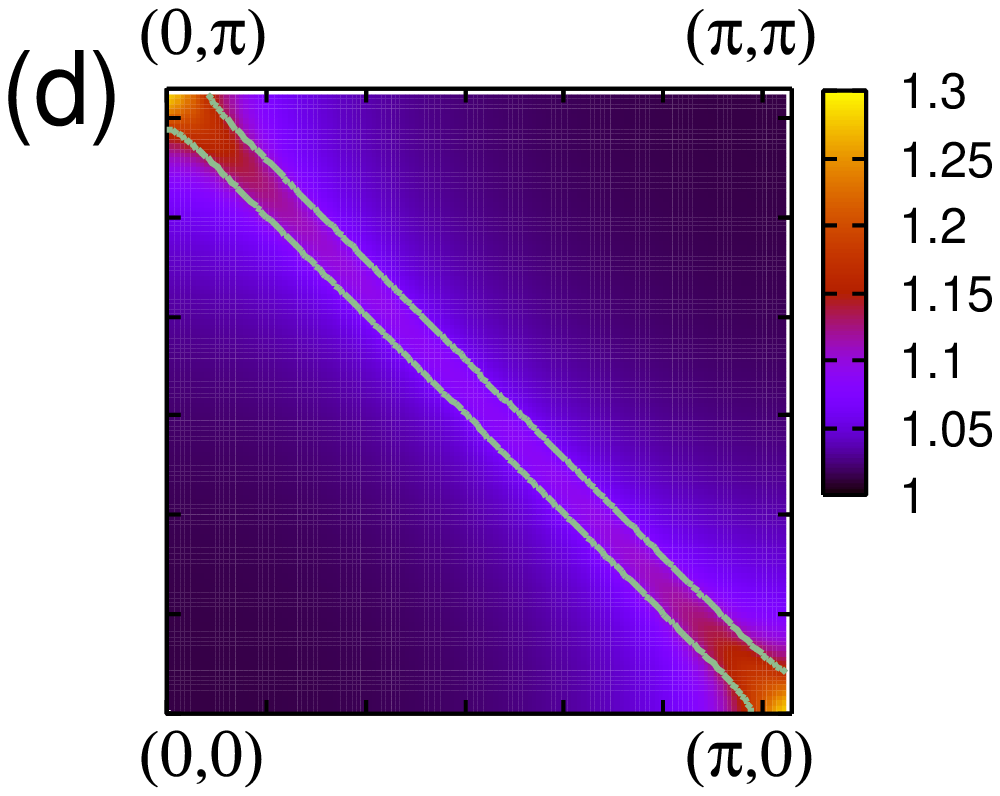}
  \end{center}
  \caption{(Color online)
    Momentum dependence of the renormalization in RSOC. 
    We show $|\mib{g}'(\mib{k},\omega=0)|/|\mib{g}(\mib{k})|$ for fillings (a) $n=0.4$, (b) 0.6, (c) 0.8, and (d) 1.0.
    Solid lines show the spin-split Fermi surfaces.
  }
  \label{Figfs}
\end{figure}

The renormalization of RSOC is anisotropic in the momentum space as shown in Fig.~\ref{Figfs}. 
Generally speaking, a large enhancement of RSOC occurs in the vicinity of the Fermi surface, 
particularly at the momentum $\mib{k} = \mib{k}_{\rm F} \parallel$ [100] and [010]. 
The spin texture indicated by the direction of the g-vector is also changed by the electron correlation effect (not shown). 
Although the correlation effect on the spin texture is negligible in the weak coupling regime, 
the spin texture may be considerably changed in the strong coupling regime.

\begin{figure}[htbp]
  \begin{center}
    \includegraphics[scale=0.35]{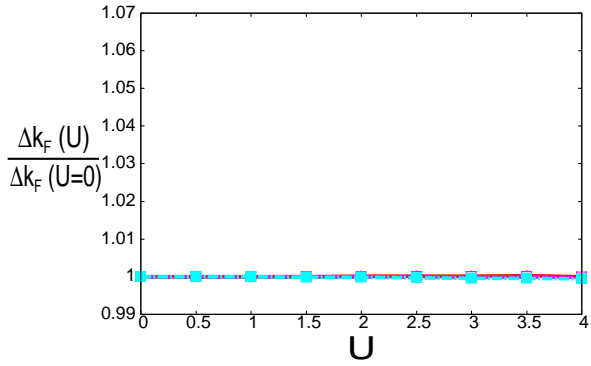}
  \end{center}
  \caption{(Color online)
    $U$-dependence of SFM $\Delta k_{\rm F}$ for $\mib{k} \parallel$ [110]. 
    We plot $\Delta k_{\rm F}$ for various electron densities $n=$0.2, 0.4, 0.6, 0.8, and 1.0, but 
    $\Delta k_{\rm F}(U) \simeq  \Delta k_{\rm F}(U=0)$ irrespective of the electron density. 
    The scale of the horizontal axis is chosen to be the same as that in Fig.~\ref{Figrg} in order to show the 
    almost complete cancellation between the $k$-mass renormalization and the enhancement of RSOC. 
  }
  \label{Figssfs}
\end{figure}

Finally, we show the SFM $\Delta k_{\rm F}$. 
Although the enhancement of renormalized RSOC implies an increase in $\Delta k_{\rm F}$, 
the SFM remains unchanged by the electron correlation effect, as shown in Fig.~\ref{Figssfs}. 
Note that the scale of the horizontal axis is chosen to be the same as in Fig.~\ref{Figrg}. 
We can not observe the electron correlation effect on $\Delta k_{\rm F}$ at this scale.  
Thus, it is indicated that the enhancement of RSOC is compensated by other effects. 
The $k$-mass renormalization indeed cancels out the enhancement of RSOC, as we show in the next subsection.

\subsection{Analytic calculation}\label{ANARES}

We here clarify the cancellation between the $k$-mass renormalization and the renormalization of RSOC. 
In this subsection, we adopt the lowest-order theory with respect to RSOC, which is justified for a weak RSOC, 
$\alpha |\mib{g}(\mib{k})| \ll \varepsilon_{\rm F}$, as realized in most non-centrosymmetric metals. 
The SFM is obtained as 
\begin{eqnarray}
\Delta k_{\rm F}\simeq\frac{2\alpha|\mib{g}'(\mib{k}_{\rm F},0)|}{|\mib{v}^{{\rm k}}(\mib{k}_{\rm F},0)|},
\label{delk}
\end{eqnarray}
where $\mib{k}_{\rm F}$ is the Fermi momentum at $\alpha=0$ and 
$\mib{v}^{{\rm k}}(\mib{k}_{\rm F},\omega) = \frac{\partial\varepsilon(\mib{k})}{\partial\mib{k}} 
+ \frac{\partial{\rm Re}\Sigma^{\rm R}_0(\mib{k},\omega)}{\partial\mib{k}}|_{\mib{k}=\mib{k}_{\rm F}}$ 
is the Fermi velocity renormalized by the $k$-mass. 
Thus, the SFM is affected by the renormalization of the $k$-mass and RSOC but not affected by the $\omega$-mass.

The perturbation expansion in terms of RSOC is carried out by expanding the noninteracting 
Green function as 
\begin{eqnarray}
&& \hspace{-8mm} G^{(0)}_{\lambda}\simeq G^{(0)} + G^{(0)} \lambda \alpha |\mib{g}|G^{(0)}
+ G^{(0)} \lambda \alpha |\mib{g}| G^{(0)} \lambda \alpha |\mib{g}|G^{(0)} + \cdots, 
\nonumber \\ &&
\end{eqnarray}
where $G^{(0)}(k)=[\omega-\varepsilon(\mib{k})]^{-1}$.
Up to the first order in $\alpha$, the diagonal self-energy $\Sigma_0(k)$ is obtained from 
the diagram in Fig.~\ref{Figse2o}(2A). It is the zeroth-order term with respect to $\alpha$. 
On the other hand, Fig.~\ref{Figse2o}(2B) represents the first-order terms 
in the off-diagonal self-energy $\Sigma_x(k)$ and $\Sigma_y(k)$. 
Thus, the lowest-order terms of the self-energy are obtained within the first order of $\alpha$. 

Differentiating the diagonal self-energy, we obtain the renormalized Fermi velocity as 
{\setlength\arraycolsep{1pt}
\begin{eqnarray}
\label{eq18}
v^{{\rm k}}_x(k)&=&v_x(\mib{k})+U^2{\rm Re}\sum_q\phi(q)G^{(0)}(q-k)^2v_x(\mib{q}-\mib{k}), \label{vkmx} \\
v^{{\rm k}}_y(k)&=&v_y(\mib{k})+U^2{\rm Re}\sum_q\phi(q)G^{(0)}(q-k)^2v_y(\mib{q}-\mib{k}), \label{vkmy}
\end{eqnarray}}
where
{\setlength\arraycolsep{3pt}
\begin{eqnarray}
\phi(q)\equiv\sum_{k'}G^{(0)}(k')G^{(0)}(q-k'), 
\label{phi}
\end{eqnarray}}
and $v_\alpha(\mib{k}) = \partial \varepsilon(\mib{k})/\partial k_\alpha$.
On the other hand, the renormalized g-vector is obtained as 
{\setlength\arraycolsep{1pt}
\begin{eqnarray}
g'_x(k)&=&g_x(\mib{k})+U^2{\rm Re}\sum_q\phi(q)G^{(0)}(q-k)^2g_x(\mib{q}-\mib{k}), \label{gpx}\\
g'_y(k)&=&g_y(\mib{k})+U^2{\rm Re}\sum_q\phi(q)G^{(0)}(q-k)^2g_y(\mib{q}-\mib{k}). \label{gpy}
\label{eq22}
\end{eqnarray}}
The derivation of Eqs.~(\ref{eq18})-(\ref{eq22}) is given in the Appendix. 
Now the similarity between the Fermi velocity and RSOC is clear. 
The Fermi velocity and RSOC acquire the same renormalization when 
the relation $\mib{g}(\mib{k})=C [-v_y(\mib{k}),v_x(\mib{k}),0]$ is satisfied with $C$ being an arbitrary constant. 
This relation is indeed satisfied in the Rashba-Hubbard model adopted in Sect.~3.1. 
Then, the SFM is not renormalized by the electron correlation effect, 
\begin{eqnarray}
\Delta k_{\rm F} \simeq 
\frac{2\alpha|\mib{g}'(\mib{k}_{\rm F},0)|}{|\mib{v}^{{\rm k}}(\mib{k}_{\rm F},0)|} 
= \frac{2\alpha|\mib{g}(\mib{k}_{\rm F})|}{|\mib{v}(\mib{k}_{\rm F})|}. 
\label{delk2}
\end{eqnarray} 
The cancellation is not complete owing to higher-order terms with respect to the RSOC $\alpha$, 
but we see almost complete cancellation between the $k$-mass renormalization and the enhanced RSOC 
for a moderate RSOC of $\alpha =0.1$ (see Fig.~\ref{Figssfs}).

\begin{figure}[htbp]
  \begin{center}
    \includegraphics[scale=0.35]{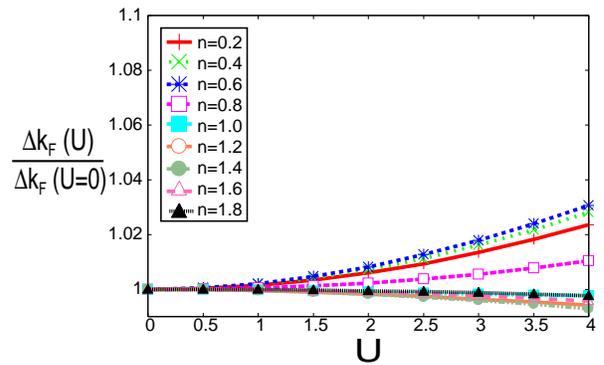}
  \end{center}
  \caption{(Color online)
    SFM $\Delta k_{\rm F}$ for $n=$0.6, 0.4, 0.2, 0.8, 1.0, 1.8, 1.6, 1.2, and 1.4 from top to bottom. 
    We assume $(t_1,t_2)=(1,0.3)$, although we assumed $(t_1,t_2)=(1,0)$ in Fig.~\ref{Figssfs}. 
    The other parameters are the same as those in Fig.~7.
  }
  \label{Fignvssfs}
\end{figure}

The above results imply that the SFM is renormalized 
by the electron correlation when the relation $\mib{g}(\mib{k})=C [-v_y(\mib{k}),v_x(\mib{k}),0]$ is not satisfied. 
For instance, we can choose the parameters $(t_1,t_2)=(1,0.3)$ so that 
$\mib{g}(\mib{k}) \not\propto [-v_y(\mib{k}),v_x(\mib{k}),0]$. 
Indeed, Fig.~\ref{Fignvssfs} shows that the SFM is affected by the Coulomb interaction. 
A finite correction to the SFM is obtained, although it 
is substantially reduced by the cancellation of the $k$-mass renormalization and the enhancement of RSOC. 
The sign of the correction depends on the band structure. We see that the SFM is enhanced (suppressed) by the 
electron correlation effect when the Fermi surface is electron-like (hole-like). 
This particle-hole asymmetry in the correction to the SFM is caused by the next-nearest-neighbour 
hopping $t_2$, which induces the particle-hole asymmetry in the band structure. 
However, we note that the robust SFM independent of $U$ (see Fig.~7 for example) does not require  
the particle-hole symmetry in the band structure. As we showed above, the SFM is not renormalized by the electron 
correlation when $\mib{g}(\mib{k})=C [-v_y(\mib{k}),v_x(\mib{k}),0]$. 
We can choose the $g$-vector so as to satisfy this relation even when $t_2 \ne 0$.

\section{Third-Order Perturbation Theory}\label{3PT}

\begin{figure}[htbp]
  \begin{center}
    \includegraphics[scale=0.5]{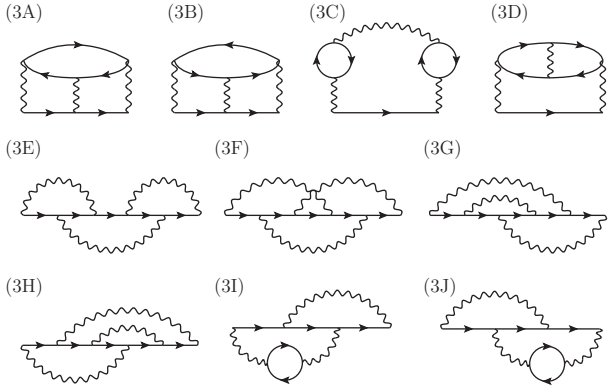}
  \end{center}
  \caption{Diagrammatic representation of third-order terms of self-energy. 
    }
  \label{Figse3o}
\end{figure}

So far we have investigated the weak coupling region of the Rashba-Hubbard model on the basis of second-order 
perturbation theory. In this section, we examine higher-order corrections by comparing 
third-order perturbation theory with second-order perturbation theory. 
The third-order terms of self-energy are diagrammatically represented in Fig.~\ref{Figse3o}.
These terms are classified according to the leading order with respect to the RSOC $\alpha$. 
The zeroth-order terms of $\alpha$ are $\Sigma^{(\rm{3A})}$ and $\Sigma^{(\rm{3B})}$, while 
$\Sigma^{(\rm{3C})}$ and $\Sigma^{(\rm{3F)-(3J})}$ are first-order terms. 
Since $\Sigma^{(\rm{3D})}$ and $\Sigma^{(\rm{3E})}$ are higher-order terms, they are negligible for $\alpha =0.1$. 

\begin{figure}[htbp]
\begin{center}
    \includegraphics[scale=0.33]{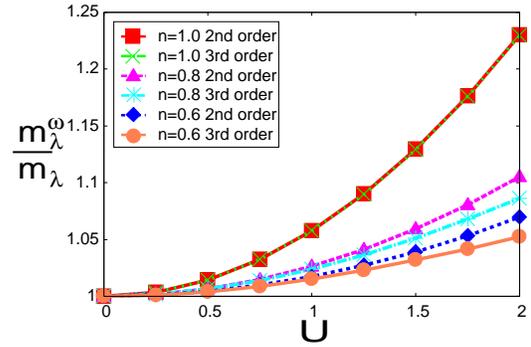}
  \end{center}
  \caption{(Color online)
    $U$-dependence of the $\omega$-mass in the $\lambda=+$ band at $\mib{k} = \mib{k}_{{\rm F} +} \parallel$ [110]. 
    Crosses, stars, and circles show the results of third-order perturbation theory 
    for $n=$1.0, 0.8, and 0.6, respectively. The results of second-order perturbation theory are shown 
    for $n=1.0$ (squares), 0.8 (triangles), and 0.6 (diamonds) for comparison. 
  }
  \label{Figp3om}
\end{figure}

\begin{figure}[htbp]
\begin{center}
    \includegraphics[scale=0.33]{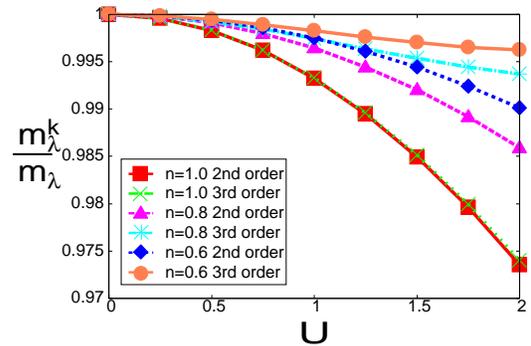}
  \end{center}
  \caption{(Color online)
    $U$-dependence of the $k$-mass in the $\lambda=+$ band at $\mib{k} = \mib{k}_{{\rm F} +} \parallel$ [110]. 
    Symbols indicate the same electron density and the same order of perturbation theory as those in Fig.~\ref{Figp3om}. 
  }
  \label{Figp3km}
\end{figure}

First, we assume the dispersion relation $(t_1,t_2)=(1,0)$, as in Sect.~3.1, and calculate the 
renormalization of the effective mass, RSOC, and SFM. 
Figures~\ref{Figp3om} and \ref{Figp3km} show the $\omega$-mass and $k$-mass on the $\lambda=+$ band, respectively.
As shown for the Hubbard model without spin-orbit coupling,~\cite{Yanase_review} the third-order correction 
partly cancels the second-order terms. Indeed, both $\omega$-mass and $k$-mass renormalization are suppressed 
by the third-order terms. The third-order terms give rise to a particularly large correction to the $k$-mass. 
A special case appears at half-filling, $n=1$. In this case, the third-order terms are negligible because 
the two leading-order terms $\Sigma^{(\rm{3A})}$ and $\Sigma^{(\rm{3B})}$ cancel each other because of 
the particle-hole symmetry~\cite{Yanase_review}.

\begin{figure}[htbp]
  \begin{center}
    \includegraphics[scale=0.35]{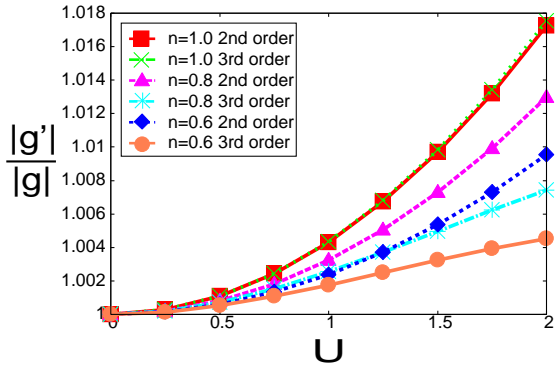}
  \end{center}
  \caption{(Color online)
    $U$-dependence of the renormalization of RSOC at $\mib{k} = \mib{k}_{{\rm F} +} \parallel$ [110]. 
    Symbols indicate the same electron density and the same order of perturbation theory as those in Fig.~\ref{Figp3om}. 
  }
  \label{Figp3rg}
\end{figure}

\begin{figure}[htbp]
  \begin{center}
    \includegraphics[scale=0.35]{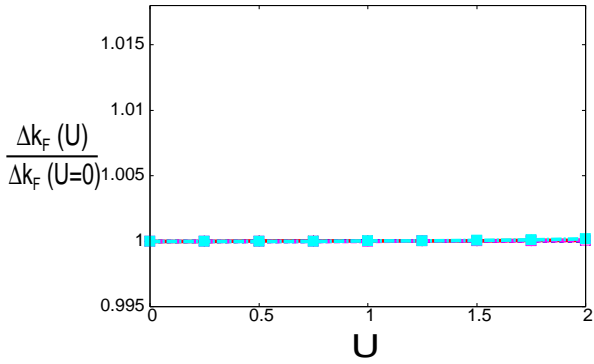}
  \end{center}
  \caption{(Color online)
    SFM obtained by third-order perturbation theory for $(t_1,t_2)=(1,0)$. 
    We show the results for electron densities $n=0.2$, 0.4, 0.6, 0.8, and 1.0. 
    The momentum is chosen to be parallel to the [110] axis. 
    The SFM is invariant against the electron correlation effect on the whole Fermi surface 
    irrespective of the electron density. 
    The scale of the horizontal axis is chosen to be the same as that in Fig.~\ref{Figp3rg}. 
  }
  \label{Figp3ssfs}
\end{figure}

The renormalization of RSOC on the Fermi surface of the $\lambda=+$ band is shown in Fig.~\ref{Figp3rg}. 
It is shown that the third-order correction reduces the enhancement of RSOC. 
Interestingly, the SFM is invariant against the electron correlation even when we take into account third-order terms 
(see Fig.~\ref{Figp3ssfs}). Thus, the cancellation between the renormalization of RSOC and the $k$-mass 
is not an artifact of second-order perturbation theory. Indeed, the cancellation occurs in each order of $U$ 
when $\mib{g}(\mib{k})=C [-v_y(\mib{k}),v_x(\mib{k}),0]$. 
Therefore, it is expected that the SFM is robust against electron correlations not only in the weak coupling regime 
but also in the strong coupling regime.

\begin{figure}[htbp]
  \begin{center}
    \includegraphics[scale=0.35]{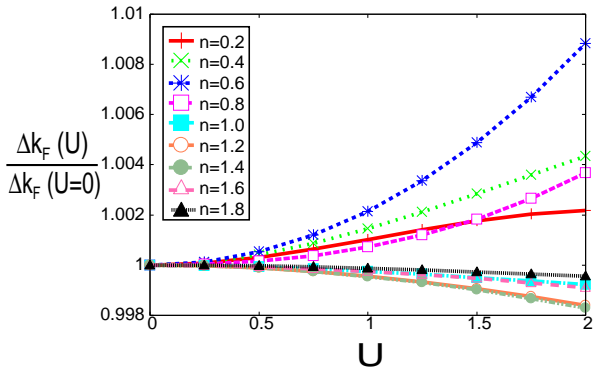}
  \end{center}
  \caption{(Color online)
    SFM obtained by third-order perturbation theory for $(t_1,t_2)=(1,0.3)$.
    $\Delta k_{\rm F}$ for $n=$0.6, 0.4, 0.8, 0.2, 1.8, 1.0, 1.6, 1.2, and 1.4 are shown 
    from top to bottom at $U=2$. 
  }
  \label{Figp3nvssfs}
\end{figure}

Next, we choose the parameters $(t_1,t_2)=(1,0.3)$, as in Fig.~\ref{Fignvssfs}, so that 
$\mib{g}(\mib{k}) \not\propto [-v_y(\mib{k}),v_x(\mib{k}),0]$. 
Then, the SFM is renormalized by the electron correlation effect as shown in Fig.~\ref{Figp3nvssfs}. 
Thus, we obtain qualitatively the same results as those in second-order perturbation theory, 
although the electron correlation effects are reduced by the third-order correction terms.

\section{Summary and Discussion}\label{SUM}

Exotic quantum phases and intriguing electromagnetic responses are induced by the antisymmetric spin-orbit coupling 
in non-centrosymmetric metals~\cite{Spin_Hall_1,Spin_Hall_2,Chiral_Magnetism_1,Chiral_Magnetism_2,NCSC}. 
Naturally, the interplay between electron correlation effects and spin-orbit coupling in non-centrosymmetric metals 
has attracted interest. 
We have investigated electron correlation effects in the two-dimensional Rashba-Hubbard model, 
which is a minimal model for a non-centrosymmetric (chiral) Fermi liquid. 
The effective mass, the renormalization of spin-orbit coupling, and the spin-split Fermi surfaces
were calculated on the basis of perturbation theory. 
We showed that the electron correlation enhances the spin-orbit coupling but the SFM  $\Delta k_{\rm F}$ 
is almost invariant against Coulomb interaction. 

In second-order perturbation theory, both numerical and analytic calculations show that 
the enhancement of spin-orbit coupling is cancelled by the $k$-mass renormalization. 
Thus, $\Delta k_{\rm F}$ is not renormalized by the electron correlation. 
The cancellation is complete when the g-vector of RSOC is represented by the velocity as  
$\mib{g}(\mib{k})=C[-v_y(\mib{k}),v_x(\mib{k}),0]$. 
Otherwise, a finite correction to the SFM appears, but it is reduced by an incomplete cancellation. 

We numerically examined the validity of second-order perturbation theory by calculating third-order correction terms 
with respect to the Coulomb interaction. Generally speaking, third-order terms partly cancel the leading-order 
second-order terms~\cite{Yanase_review}. 
Indeed, the renormalization of the effective mass and spin-orbit coupling is decreased by the third-order terms. 
The SFM remains invariant against the electron correlation in third-order perturbation theory 
when the relation $\mib{g}(\mib{k})=C[-v_y(\mib{k}),v_x(\mib{k}),0]$ is satisfied. 
Thus, the robustness of the SFM against the electron correlation is not an artifact of second-order perturbation theory 
and is expected to be an exact property.

Although we considered a Rashba-type spin-orbit coupling, 
our results are generally valid for other kinds of antisymmetric spin-orbit coupling. 
For example, we confirmed that the spin-orbit coupling is enhanced but the SFM is invariant against 
the electron correlation effect when the g-vector is described as $\mib{g}(\mib{k})=C[v_x(\mib{k}),v_y(\mib{k}),0]$. 
This antisymmetric spin-orbit coupling is allowed in crystals having   
$D_n$, $C_1$, or $C_2$ point group symmetry, such as the non-centrosymmetric superconductor 
UIr.~\cite{doi:10.1143/JPSJ.73.3129}  
Generally speaking, the SFM is not renormalized by the electron correlation when the g-vector of antisymmetric 
spin-orbit coupling is linearly related to the velocity of quasiparticles. 
This condition is satisfied in many theoretical models adopted for non-centrosymmetric systems. 
For instance, two-dimensional electron gases formed on semiconductor heterostructures~\cite{Winkler_book} and 
oxide interfaces~\cite{Ohtomo,Ueno_review} have been studied on the basis of the model with 
the isotropic dispersion relation $\varepsilon(\mib{k}) = \mib{k}^2/2m$ 
and RSOC with $\mib{g}(\mib{k})=(-k_y, k_x,0)$~\cite{Maslov2013,Raikh1999,Ross2005,Chesi2011,Chesi2012,
Michaeli,Agterberg_skyrmion}. 
Then, we indeed see the relation $\mib{g}(\mib{k})=m[-v_y(\mib{k}),v_x(\mib{k}),0]$. 
Cold atom gases with a tunable synthetic spin-orbit coupling~\cite{Spielman} also satisfy the condition. 
One-, two-, and three-dimensional Rashba-Hubbard models satisfying the condition have been 
studied~\cite{Yanase_CePt3Si,JPSJ.77.124711,Yokoyama,PhysRevB.80.140509,cond-mat.1406.7293}. 
On the other hand, crystals having a $T_{\rm d}$ point group symmetry do not satisfy the condition 
because the g-vector of Dresselhaus-type spin-orbit coupling~\cite{Dresselhaus} 
is represented by cubic terms with respect to the momentum near the $\Gamma$ point of the Brillouin zone.

The relation of importance, $\mib{g}(\mib{k}) = C[-v_y(\mib{k}),v_x(\mib{k}),0]$, is also broken 
in orbitally degenerate systems. 
According to the derivation of RSOC based on multiorbital 
models~\cite{JPSJ.77.124711,doi:10.7566/JPSJ.82.044711,doi:10.7566/JPSJ.82.083705,zhong2013theory}, 
the relation is approximately satisfied when the orbital degeneracy 
is substantially lifted by a large crystal electric field. 
Otherwise, both the quasiparticle velocity and the g-vector of RSOC acquire a complicated momentum dependence, 
and thus they do not show a linear relation~\cite{doi:10.7566/JPSJ.82.044711,doi:10.7566/JPSJ.82.083705}.  
For example, a cubic RSOC has been observed in the two-dimensional electron gas on a SrTiO$_3$ surface~\cite{Cubic_Rashba}, 
and it may be induced by the orbital degree of freedom in $t_{2g}$ electrons~\cite{zhong2013theory}. 
Then, the amplitude of the SFM as well as the spin texture in the momentum space should be affected 
by the electron correlation. 
Thus, analyses of the multiorbital Rashba-Hubbard model and periodic Rashba-Anderson model 
are of particular interest, and therefore, we will study them in the near future. 
It is expected that the spin-orbit coupling will be renormalized through the renormalization of the crystal electric field, 
as seen in a GW calculation based on density functional theory~\cite{Rusinov}.

Finally, we comment on {\it locally} non-centrosymmetric metals, which have global inversion symmetry 
but lack a local inversion symmetry on atoms. 
Instead of the uniform antisymmetric spin-orbit coupling discussed in this paper, a staggered antisymmetric 
spin-orbit coupling gives rise to a magneto-electric effect~\cite{doi:10.7566/JPSJ.83.014703} 
and exotic superconductivity~\cite{Yoshida_1,doi:10.7566/JPSJ.83.061014}.  
It is expected that the staggered spin-orbit coupling will be enhanced by the electron correlation, 
as we found for non-centrosymmetric metals. 
Since the contribution of the staggered spin-orbit coupling is determined by comparison with 
the inter-sublattice hopping~\cite{Maruyama_1}, it is an important future issue to calculate their renormalization.

After we submitted the first manuscript, we became aware of a theoretical work~\cite{Fujimoto2015} 
on a related subject.
Reference~58 showed a substantial deformation of spin-split Fermi surfaces and the SFM near the magnetic quantum critical point, 
in sharp contrast to our results. We believe that the discrepancy is (at least partly) owing to the fact that 
the $k$-mass renormalization, which plays an essential role in our conclusion, is neglected in Ref.~58.

\section*{Acknowledgements}
The authors are grateful to T. Yoshida for valuable discussions and to D. L. Maslov for helpful comments. 
This work was supported by the ``Topological Quantum Phenomena" (No. 25103711) KAKENHI on Innovative Areas 
from MEXT of Japan and by JSPS KAKENHI Grant Numbers 24740230 and 15K05164.

\appendix
\section{Fermi velocity and RSOC renormalized by electron correlation}

We here derive the Fermi velocity and RSOC renormalized by the electron correlation effects 
in second-order perturbation theory. 
We assume $\alpha |\mib{g}(\mib{k})| \ll \varepsilon_{\rm F}$, as realized in most non-centrosymmetric metals. 
Then, the lowest-order theory with respect to the coupling constant of RSOC is justified.

The Fermi velocity renormalized by the $k$-mass $\mib{v}^{\rm k}(k)$ is obtained in the zeroth order of $\alpha$ as 
{\setlength\arraycolsep{3pt}
\begin{eqnarray}
\mib{v}^{{\rm k}}(k) &=& \frac{\partial\varepsilon(\mib{k})}{\partial\mib{k}}
+\frac{\partial{\rm Re}\Sigma_0^{\rm R}(k)}{\partial\mib{k}}
+\lambda\alpha\frac{\partial|\mib{g}'(k)|}{\partial\mib{k}} \nonumber \\ 
&\rightarrow&\frac{\partial\varepsilon(\mib{k})}{\partial\mib{k}}
+\frac{\partial{\rm Re}\Sigma_0^{\rm R}(k)}{\partial\mib{k}} \biggl{|}_{\alpha=0}. 
%\ \ \ \ \ (\alpha\rightarrow 0).
\end{eqnarray}}
At $\alpha =0$, the diagonal self-energy is obtained by the Feynman diagram in Fig.~\ref{Figse2o}(2A), 
and thus $\Sigma_0(k) = \Sigma_{\uparrow\uparrow}^{(2{\rm A})}(k)$. 
Therefore, we obtain the Fermi velocity renormalized by the $k$-mass as, 
{\setlength\arraycolsep{3pt}
\begin{eqnarray}
\mib{v}^{{\rm k}}(k) &\simeq&
\frac{\partial\varepsilon(\mib{k})}{\partial\mib{k}}
+\frac{\partial{\rm Re}\Sigma_{\uparrow\uparrow}^{{\rm R}(2{\rm A})}(k)}{\partial\mib{k}} \bigg{|}_{\alpha =0}. 
\label{A2}
\end{eqnarray}}
The $x$- and $y$-components of Eq.~(\ref{A2}) are expressed by Eqs.~(\ref{vkmx}) and (\ref{vkmy}), respectively.

The renormalization of RSOC is obtained by calculating the renormalized g-vector $\mib{g}'(k)$, 
which we obtain in the limit $\alpha \rightarrow 0$ as 
%{\setlength\arraycolsep{3pt}
\begin{eqnarray}
\hspace{-12mm}
&&\mib{g}'(k) = \mib{g}(\mib{k}) + \frac{{\rm Re}\mib{\Sigma}^{\rm R}(k)}{\alpha}
\rightarrow
\mib{g}(\mib{k})+\frac{\partial{\rm Re}\mib{\Sigma}^{\rm R}(k)}{\partial\alpha} \bigg{|}_{\alpha =0}. 
\label{A3}
\end{eqnarray}
%}
The spin-dependent self-energy $\mib{\Sigma}^{\rm R}$ is obtained up to the first order of $\alpha$, 
%{\setlength\arraycolsep{3pt}
\begin{eqnarray}
\mib{\Sigma}(k) = \mib{\Sigma}^{(2{\rm B})}(k). 
\label{A4}
\end{eqnarray}
%}
Thus, the renormalized g-vector is obtained by calculating the off-diagonal self-energy represented 
by the Feynman diagram in Fig.~\ref{Figse2o}(2B). 
Differentiating the self-energy with respect to $\alpha$, 
we obtain Eqs.~(\ref{gpx}) and (\ref{gpy}) from Eq.~(\ref{A3}).

%\bibliographystyle{jpsj}
%\bibliography{ncs}

\begin{thebibliography}{10}
\bibitem{comment1}
The spin-orbit coupling is also renormalized by the $\omega$-mass. 
However, the $\omega$-mass renormalization reduces the kinetic energy 
as well as the spin-orbit coupling, and thus the ratio between them is not 
affected. Therefore, we define the renormalized RSOC as $\alpha \mib{g}'(k) 
\equiv \alpha \mib{g}(\mib{k}) + {\rm Re} \mib{\Sigma}^{\rm R}(k)$.
\end{thebibliography}

\end{document}